\documentclass[aps,prb,twocolumn,showpacs,amsmath,amssymb,groupedaddress]{revtex4-1}
\usepackage{graphicx}
\usepackage{dcolumn}
\usepackage{bm}
\usepackage{color}
\usepackage{longtable}
\usepackage{multirow}

\begin{document}
\date{\today}

\title{Direct Observation of an Incommensurate Charge Density Wave in the BiS$_2$-based Superconductor NdO$_{1-x}$F$_x$BiS$_2$}

\author{Jooseop Lee$^{1,2}$}
\author{Masanori Nagao$^{3}$} 
\author{Yoshikazu Mizuguchi$^{4}$} 
\author{Jacob Ruff$^{2}$} 

\affiliation{$^{1}$CALDES, Institute for Basic Science, Pohang 37673, Republic of Korea}
\affiliation{$^{2}$CHESS, Cornell University, Ithaca, New York 14853, USA}
\affiliation{$^{3}$University of Yamanashi, Kofu, Yamanashi 400-8511, Japan}
\affiliation{$^{4}$Department of Physics, Tokyo Metropolitan University, Hachioji, Tokyo 192-0397, Japan}

\begin{abstract}

The nature of superconductivity in BiS$_2$-based superconductors has been controversial while ab-initio calculations proposed this system in close proximity to a charge-density-wave (CDW) phase. Using high-energy high-flux X-ray diffraction, we reveal an intrinsic and long-range CDW phase coexisting with superconductivity in NdO$_{1-x}$F$_{x}$BiS$_2$ superconductor ($x$ = 0.37 and 0.3). The CDW wavevector in NdO$_{0.63}$F$_{0.37}$BiS$_2$ correspond \textbf{Q}$_{\rm{CDW}}$ = (0.17, 0.17, 0.5) and is associated with transverse atomic displacements. Interestingly, this wavevector does not match theoretical expectations based on either phonon softening or Fermi surface nesting. In NdO$_{0.7}$F$_{0.3}$BiS$_2$, where the superconducting transition temperature is highest, the CDW satellites are slightly broader and weaker compared to NdO$_{0.63}$F$_{0.37}$BiS$_2$, possibly suggesting the competition with the superconductivity. Lastly, we measure a thermal diffuse scattering across the superconducting transition temperature and find no meaningful changes. Our result suggests the importance of understanding CDW which might hold a key to the superconductivity in the BiS$_2$-based superconductor.
% in favor of the unconventional pairing mechanism

\end{abstract}

\maketitle

\section{INTRODUCTION}

The BiS$_2$-based superconductors have many similarities with the unconventional cuprates and iron-based superconductors \cite{First, MizuRev, DeguRev, YaziciRev, UsuiRev, TeraRev}. Their layered structure is low dimensional composed of BiS$_2$ superconducting layers and Ln$_2$O$_2$ (Ln: Lanthanide) blocking layers alternating along the c-axis. Moreover, Bi ions responsible for the superconductivity are arranged in a square lattice. On the other hand, Bi is not magnetic and shows 6p$_{x,y}$ orbital characters, which have less electron correlation compared with 3d orbitals \cite{UsuiPRB}. The highest transition temperature reported is relatively low $\approx$ 11.5 K \cite{HighestTc}.

The nature of superconductivity in BiS$_2$-based superconductors, therefore, has been controversial whether it can be considered as a conventional phonon-mediated BCS superconductor or another family of non-BCS superconductors that opens a new route to unconventional superconductivity. Density functional theory (DFT) calculations have predicted a significant electron-phonon (e-ph) coupling \cite{Yildirim, WanPRB, LiEPL, FengJAP}, that was later supported by Raman experiment \cite{RamanPRB}. Meanwhile, many ab initio studies also find strong Fermi surface nesting (FSN) near optimal doping \cite{Yildirim, WanPRB, UsuiPRB} that can lead to unconventional pairing mechanisms \cite{T2, T8, T9, T10, T15, T28, T29}. Inelastic neutron scattering on polycrystalline samples \cite{JLEE2013} and Raman studies \cite{RamanSST} indicate that the possible e-ph coupling can be much weaker than theoretically expected. Yet, a momentum-resolved phonon study on a single-crystalline sample is still lacking.

Interestingly, theoretical studies predict negative energy phonon modes possibly related to the FSN that leads to ferroelectric or charge-density-wave (CDW) phases \cite{Yildirim, WanPRB}. According to the ab-inito calculation, in LaOBiS$_2$, there is a ferroelectric soft phonon at the zone center, whose shallow potential curve prevents the development of static long-range order. In the optimally doped LaO$_{0.5}$F$_{0.5}$BiS$_2$, on the other hand, unstable phonon branches are expected along the $\bm{q}$ = ($\delta$, $\delta$, 0) causing the in-plane distortion of the BiS$_2$ plane, especially the shift of S atoms by about 0.16 $\rm{\AA}$ away from the high-symmetry position. Total energy analysis finds the structure with the lowest energy to be a $\sqrt{2}$ x $2\sqrt{2}$ CDW superstructure with sinusoidally distorted one-dimensional Bi-S chains. 

The relation between CDW and superconductivity has been hotly debated as CDW turns out to be a ubiquitous ground state found in the phase diagram of many unconventional superconductors, for example, cuprate or transition-metal dichalcogenides (TMDC) superconductors \cite{Dean, Kis}. In BiS$_2$-based superconductors, Scanning Tunneling Microscopy (STM) studies asserted a modulation in the real-space electronic features that are characterized, for example, by a checkerboard stripe \cite{CheckerboardSTM, STM2017, STM2018}. These results, however, poses a question if the observed CDW is an intrinsic bulk property especially since they also found a large spectroscopic gap of about 40 meV contrary to a bulk metallic nature \cite{CheckerboardSTM}. 

Moreover, it has been known from the early stage that the superconducting transition temperature in BiS$_2$-based superconductors can be greatly enhanced under pressure or by pressure annealing \cite{HighestTc, HPEnhance}. First-principles energy calculations show that there is a family of polytypes nearly-degenerate in energy within $\approx$ 1 meV \cite{PolyDFT, PolyARPES} and x-ray diffraction measurement found different structures coexisting in the same batch of crystals \cite{PolyXRD}. In-plane dynamic charge fluctuations and local distortions have been claimed from neutron and x-ray diffraction and the pair density function analysis \cite{Anushika19, AnushikaNd17, AnushikaLa17, Anushika15}.  These findings all point to a crucial role of structural instability in understanding BiS$_2$-based systems.

Here, we report a long-range CDW ground state that coexists with superconductivity in NdO$_{0.63}$F$_{0.37}$BiS$_2$ observed using high-energy and high-flux x-ray diffraction. The CDW phase onsets around $\approx$ 120 K with a wavevector \textbf{Q}$_{\rm{CDW}}$ = (0.17, 0.17, 0.5). This wavevector is distinct from the values proposed from either phonon calculation or Fermi surface evaluation and calls attention to the interlayer couplings neglected in the previous studies. A detailed study on the x-ray reflection pattern shows that the CDW satellite peaks are stronger along one of the diagonal directions in the reciprocal (H, K, 0.5) plane, suggesting one-dimensional and diagonal atomic modulation that breaks 4-fold symmetry. In the NdO$_{0.7}$F$_{0.3}$BiS$_2$ with optimal electron doping \cite{NdDiscovery}, while a similar satellite diffraction pattern is observed, the CDW peaks are broader and weaker suggesting the competition between superconductivity and charge density wave in this system. Also, the relationship between phonon and superconductivity is investigated by x-ray diffuse scattering measurement. No changes in the phonon frequency across the superconducting transition temperature, T$_c$ $\approx$ 5 K, are observed. %This implies that the role of phonon may not be as significant as expected and supports unconventional pairing scenarios.

\section{METHOD}

\subsection{SAMPLE PREPARATION}

The single-crystal samples of NdO$_{1-x}$F$_x$BiS$_2$ (x = 0.3 and 0.37) were prepared as described in the Ref \cite{First, Sample2012, Sample2013}. Samples were synthesized using high temperature CsCl/KCl flux method in a vacuum-sealed quartz tube with the starting materials of Nd$_2$S$_3$, Bi, Bi$_2$S$_3$, Bi$_2$O$_3$, BiF$_3$, CsCl, and KCl at approximately 616 $^{\circ}$C. The single crystals are flat and plate-like with the c-axis perpendicular to the growth direction which can be easily exfoliated due to a Van der Waals gap between the BiS$_2$ planes \cite{CheckerboardSTM}. The typical sample size is about 2 mm x 2 mm x 0.05 mm. From resistivity measurements, the transition temperature was determined to be 5.1 K for x = 0.3 and 5.0 K for x = 0.37. The compositions were determined by electron probe microanalysis (EPMA). 

\subsection{X-RAY DIFFRACTION}

We used a high-energy and high-flux synchrotron x-ray at the A2 beamline at Cornell High Energy Synchrotron Source (CHESS). For the x = 0.3 sample, incident energy, E$_i$, of 41.75 keV and Pilatus 300 area detector were used, and the sample was cooled with a closed-cycle refrigerator. For the x = 0.37 sample, an E$_i$ = 39.80 keV was used with other conditions identical. The beam was slit down to a typical size of 100 x 100 $\mu$m. All measurements were performed in transmission geometry taking advantage of the high flux at high incident energies at CHESS. Typical reciprocal volume maps are obtained by combining two sets of 800 images at different two-theta angles with each image having an increment of theta of 0.025$^{\circ}$ and counting time 10 seconds.

\section{RESULTS}

%====================================================================
\begin{figure*}[th]
\includegraphics[width=0.95\hsize]{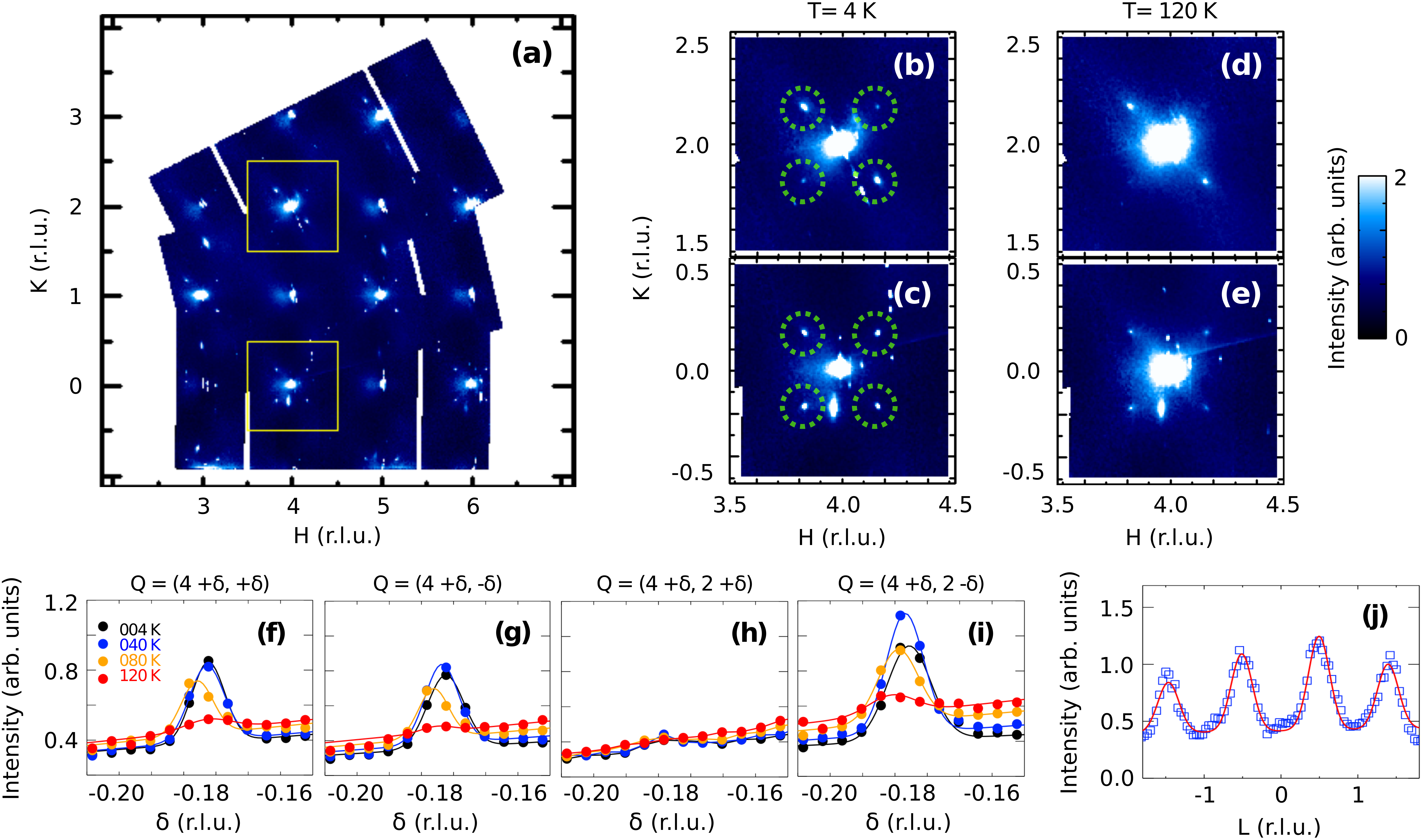}
\centering
\caption{Single-crystal x-ray diffraction of NdO$_{0.63}$F$_{0.37}$BiS$_2$ obtained at A2, CHESS with E$_i$ = 39.80 keV. (a) (H, K) reciprocal space map at L = 0.5 measured at the base temperature of 4 K. Miller indices are all in the reciprocal lattice unit (r.l.u.). Expanded view of satellite reflections at 4 K around (b) $\bm{q}$ = (4, 2) and (c) $\bm{q}$ = (4, 0), and the corresponding regions have been marked by yellow squares in (a).  Same figures at an elevated temperature of 120 K, i.e., around (d) $\bm{q}$ = (4, 2) and (e) $\bm{q}$ = (4, 0). Intensity curves measured at various temperature along (f) ($\delta$,  $\delta$)- and (g) ($\delta$,  -$\delta$)- directions at $\bm{q}$ = (4, 0) and along (h) ($\delta$,  $\delta$)- and (i) ($\delta$,  -$\delta$)- directions at $\bm{q}$ = (4, 2) showing the effect of unidirectional atomic displacement in diagonal directions. (j) L-dependence of the CDW satellite peaks at (H, K) = (4-0.17, 0.17). For all 1D curves, fits to Gaussian are shown in sold lines except (h) where there is no peak. In fitting the L-dependence, the width of all Gaussian peaks were tied to a same value.}
\label{fig:1}
\end{figure*}
%====================================================================

A high-energy single-crystal x-ray diffraction results of NdO$_{0.63}$F$_{0.37}$BiS$_2$ are shown in Fig. 1. Figure 1 (a) shows the intensity map of the reciprocal space plane (H, K) at L = 0.5 measured at 4 K. While weak in intensity, we can clearly observe sharp satellite peaks away from the Bragg peak positions with $\delta\bm{q}$ = (0.17, 0.17). Fig. 1 (b) and (c) shows a magnified view around $\bm{q}$ = (4, 2) and (4, 0), respectively, for instance. The Bragg reflection at the integer position is surrounded by four satellite reflections which are marked with dashed green circles to make it clear. These satellite peaks indicate a periodic lattice modulation that most likely results from a CDW. 

It is intriguing to note that there are lines of diffuse scattering in diagonal directions, reminiscent of the unstable phonon modes along the entire $\Gamma$-M direction suggested by ab-initio calculations \cite{Yildirim}. See the zoom-in views Fig. 1 (d) and (e) again around $\bm{q}$ = (4, 2) and (4, 0), respectively, measured at 120 K where the thermal diffuse scattering intensity is stronger due to the Bose-factor. The lines of diffuse scattering seem to be enhanced on the line from $\Gamma$ to satellite peaks. While the unstable phonon modes are merely an artifact from the calculation, the soft phonon branch along the entire line should have an important physical origin \cite{Yildirim}, for example, an extended collapse of phonon modes due to electron-phonon coupling that condenses into CDW at low temperature as in TMDC \cite{Weber}. Further energy-resolved phonon dispersion measurement will be needed to reach a firm conclusion on the origin of the thermal diffuse scattering line.

Moreover, we notice differences in the intensities of the satellite peaks. For instance, as shown in Fig. 1 (b), satellite reflections around $\bm{q}$ = (4, 2) are much stronger along ($\delta$,  -$\delta$) direction than along the other diagonal ($\delta$,  $\delta$) direction. Yet, the CDW peaks around (4, 0) all have comparable intensities, as in Fig. 1 (c). 

The intensity of CDW satellite reflections can be expressed as \cite{Overhauser, JLEE2020}
%====================================================================
\begin{equation}
    I(\bm{q}) \propto \Big|\sum_{\alpha} (\bm{q} \cdot \bm{\epsilon}_{\alpha}) f_{\alpha}(\bm{q} \pm \bm{Q}_{\rm{CDW}}) \Big|^2 \delta(\bm{q}-\bm{G} \pm \bm{Q}_{\rm{CDW}}).
\label{eq:eq1}
\end{equation}
%====================================================================
Here, $\bm{\epsilon}_{\alpha}$ and $f_{\alpha}$ are the atomic displacement and the structure factor of atom $\alpha$, respectively, while $\bm{G}$ is the reciprocal lattice vector. Considering the $(\bm{q}\cdot\bm{\epsilon}_{\alpha})$ prefactor, we conclude that the displacement of atom responsible for the CDW is almost transverse and along diagonal direction: along ($\delta$,  -$\delta$) direction for \textbf{Q}$_{\rm{CDW}}$ = (0.17, 0.17) and along ($\delta$,  $\delta$) direction for \textbf{Q}$_{\rm{CDW}}$ = (0.17, -0.17). The diffraction pattern observed should be understood as an overlay of two domains, breaking 4-fold symmetry. This unidirectional modulation is consistent with the quasi-1D nature expected from small mixing between 6p$_x$ and 6p$_y$ orbitals of Bi ions \cite{UsuiPRB}. 

Figure 1 (f)-(i) show the intensity curves at various temperatures along diagonal directions around two different Bragg peaks: $\bm{q}$ = (4, 0) for Fig. 1 (f) and (g) and $\bm{q}$ = (4, 2) for Fig. 1 (h) and (i). At all temperatures, the satellite reflections in Fig. 1 (h) has negligible intensities compared with Fig. 1 (i) due to near-orthogonality between $\bm{q}$ and $\bm{\epsilon}_{\alpha}$ while the CDW reflections around (4, 0) all have similar intensities, Fig. 1 (f) and (g). From the temperature dependence of the CDW peaks showing a rapid decrease around 120 K, we expect the onset temperature should be a bit above, but close to 120 K. Future study on the detailed temperature dependence will be required to pin down the CDW onset temperature and the order of phase transition.

A CDW phase has been conjectured from previous STM studies on BiS$_2$-based superconductors \cite{CheckerboardSTM, STM2017, STM2018}, though there have been concerns about possible surface effects and/or sample inhomogeneities. Our x-ray data, from the sharp width of satellite peaks, demonstrate that the CDW is indeed a long-range order. Furthermore, the transmission geometry ensures that the CDW is a bulk intrinsic property, rather than a surface reconstruction. The length of the wavevector, 0.172$\sqrt{2}a^{*}$, corresponds to about 3.97$a$ (=2.81$\sqrt{2}a$) in real space, consistent with STM results reporting 3$a$ $\sim$ 6$a$ in-plane superstructure \cite{CheckerboardSTM, STM2017, STM2018}. 

%====================================================================
\begin{figure*}[htp]
\includegraphics[width=0.95\hsize]{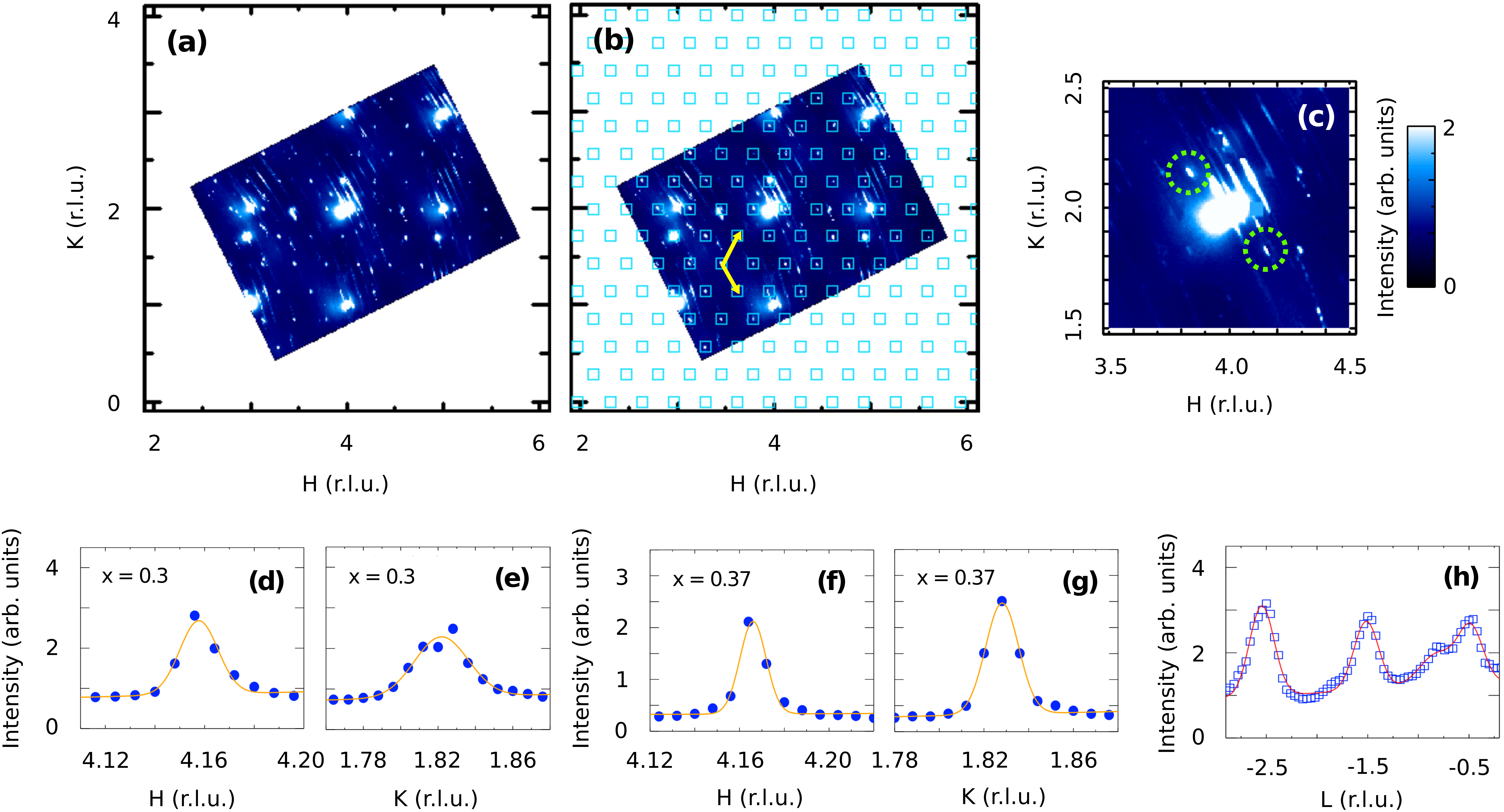}
\centering
\caption{(a) NdO$_{0.7}$F$_{0.3}$BiS$_2$ (H, K, 0.5) reciprocal space map measured at 4.3 K (b) The same map where the positions of reflections from the 3 x 3 unit cell indexed with blue squares. The corresponding reciprocal basis vectors are shown in yellow arrows. (c) A zoom-in view around $\bm{q}$ = (4, 2) which shows the CDW satellite peaks marked with green dashed circles. The contribution from the inhomogeneous phase has been removed for clarity. Intensity curves of the CDW satellite $\bm{q}$ = (4 +$\delta$, -$\delta$) (d) along H and (e) along K for NdO$_{0.7}$F$_{0.3}$BiS$_2$. For comparison, similar curves are shown (f) along H and (g) along K for NdO$_{0.63}$F$_{0.37}$BiS$_2$. (h) L-dependence at $\bm{q}$ = (4 +$\delta$, -$\delta$). For all 1D curves, fits to Gaussian are shown in sold lines. In fitting the L-dependence, the widths of all Gaussian peaks were tied to the same value and an extra Gaussian was added to account for a hump, possibly from the imperfect crystallinity.}
\label{fig:2}
\end{figure*}
%====================================================================

To check the L-dependence of the CDW satellite peaks, we made an intensity plot along L-direction at (H, K) = (4 - 0.17, 0.17) as shown in Fig. 1 (j). Surprisingly, the intensity is peaked at every half-integer position indicating doubling of the unit cell along the c-axis. The width of the peak is fitted with 0.33(1) r.l.u. corresponding to a correlation length of $\approx$ 3 unit cell. Though the broad width in Q-space suggests a weak and short-ranged interlayer correlation, the peak structure is still well-defined. It is worthy to note that previous phonon studies focused on the instabilities along $\textbf{q}$  = ($\delta$, $\delta$, 0) direction leading to a CDW of 2$\sqrt{2}$ x $\sqrt{2}$ x 1 superstructure \cite{Yildirim, WanPRB}. DFT calculation also pointed to FSN along the same direction assuming the 2D character of the Fermi surface \cite{Yildirim, WanPRB, UsuiPRB}. Our data, however, reveals that there is a clear L-dependence in the CDW reflections, indicating interlayer couplings in NdO$_{0.63}$F$_{0.37}$BiS$_2$ superconductors, which places an important constraint on future studies on the phonon, CDW, and superconductivity in BiS$_2$-based superconductors.

We repeated a similar x-ray diffraction measurement on NdO$_{0.7}$F$_{0.3}$BiS$_2$ sample to figure out the doping dependence of CDW. Figure 2 (a) shows the scattering intensity map of (H, K) plane at L = 0.5 measured at 4.3 K. At a first glance, we find unexpected numerous weak reflections at non-integer positions. We notice that the positions of these reflections are not correlated with the main Bragg peaks as expected for a CDW, and therefore cannot be from a superstructure formation. It turns out that they are indexed with a reciprocal basis vectors $\bm{a}_{1}^{*} = (\cos(60^{\circ}),~\sin(60^{\circ}))\bm{a}^{*}/3$ and $\bm{a}_{2}^{*} = (\cos(60^{\circ}),~\sin(-60^{\circ}))\bm{a}^{*}/3$ ($\bm{a}^{*}$: the original a-axis reciprocal lattice vector), which are marked with blue squares in Fig. 2 (b). This new phase must be closely linked to the original structure considering that the lattice parameters are exactly 3 times the original ones and have a precise 60-degree angle with respect to the $\bm{a}^{*}$. The hexagonal (or trigonal) reflections disappear when we decrease the beam size to tens of microns and scan other surface areas. It, therefore, might be induced from inhomogeneity in the sample, for example, by uneven doping concentration, defect, vacancy, local strain, etc. Rough estimation reveals that the volume fraction of the 3 $\times$ 3 phase is about 0.71 $\pm$ 0.08 $\%$, negligible enough to ensure that the composition is close to the nominal average one \cite{SuppleVol}. The L-dependence is not obvious in this sample due to the imperfect crystallinity along the c-axis. 

Ruling out the contribution from the 3 x 3 phase, we find the satellite reflections from the CDW as in Fig. 2 (c) around $\bm{q}$ = (4, 2). Again, the polarization dependence of the CDW satellites is clear and the atomic displacements should be unidirectional and along diagonal directions. The L-dependence is also peaked at half-integer positions consistent with x = 0.37 system.

Furthermore, we notice that the CDW satellite peaks are slightly broader and weaker compared to those of NdO$_{0.63}$F$_{0.37}$BiS$_2$. Figure 2 (d) and (e) show cuts along the H and K direction, respectively. The FWHM widths estimated from Gaussian fits are 0.018(2) r.l.u. along H and  0.034(2) r.l.u. along K. These values can be compared with those from x = 0.37 sample, 0.014(1) r.l.u. and 0.017(1) r.l.u. along H and K as shown in Fig. 2 (f) and (g), respectively. The intensity ratio with respect to the nearby Bragg peak, $\bm{q}$ = (4, 2), is 1/1597(197) compared to the ratio of 1/869(91) for x = 0.37. We estimated the domain size of the CDW \cite{Domain1, Domain2} and obtained 181(8) $\AA$ for x = 0.37 and 104(9) $\AA$ for x = 0.3 \cite{SuppleCorr}. The competition of CDW and SC, demonstrated by shorter CDW domain size and weaker reflections for higher superconducting transition temperature, is very similar to what is observed in cuprates \cite{Cuprate1, Cuprate2}. Considering that x = 0.3 is the optimal doping level with the highest T$_{c}$ among NdO$_{1-x}$F$_{x}$BiS$_2$\cite{NdDiscovery}, there could be possible competition between the superconductivity and charge density wave, though they coexist in current doping range.

%====================================================================
\begin{figure}[t]
\includegraphics[width=0.95\hsize]{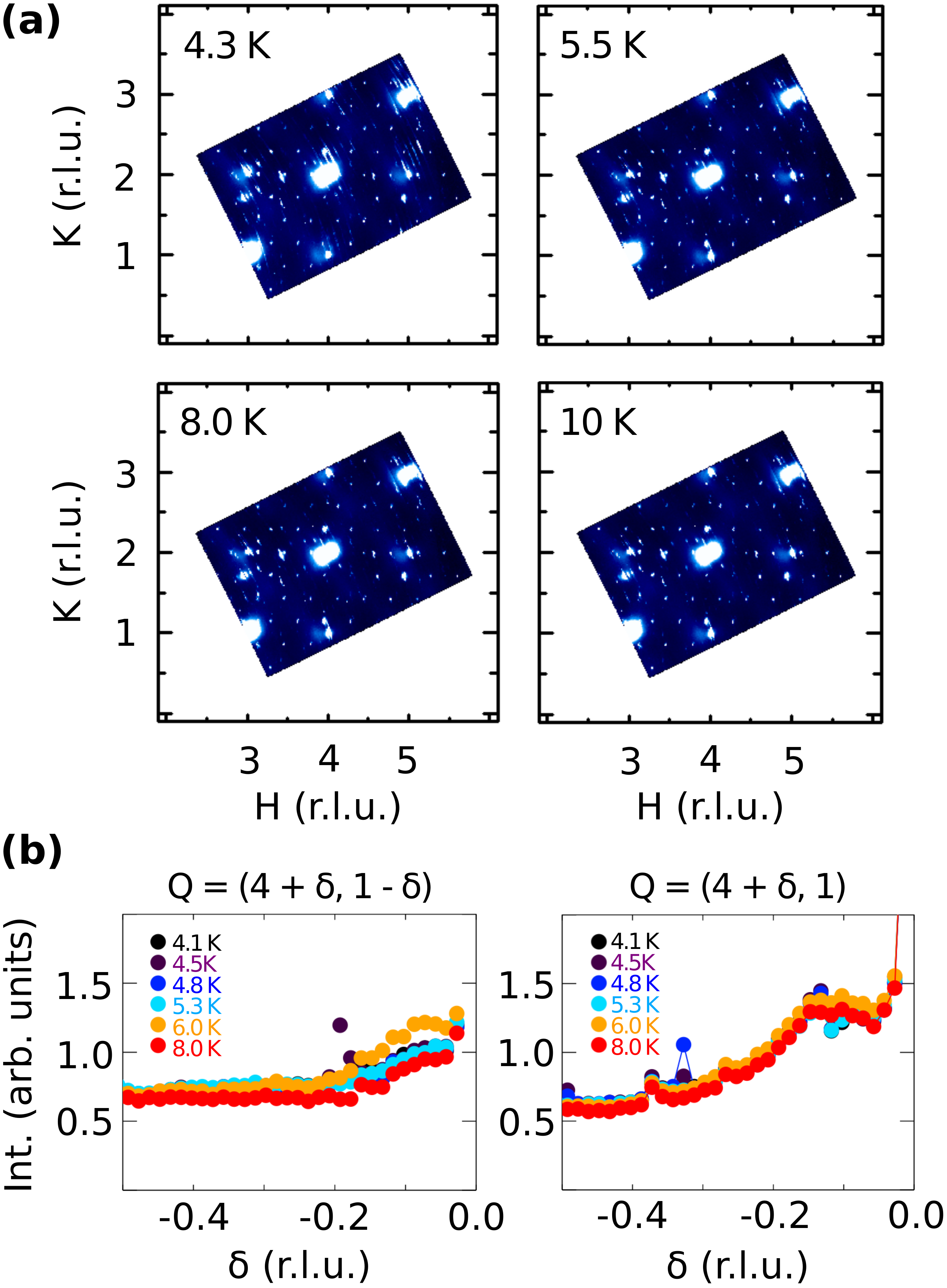}
\centering
\caption{(a) Thermal diffuse scattering intensity map of NdO$_{0.7}$F$_{0.3}$BiS$_2$ at (H, K, 0) plane across the T$_c$. (b) Intensity curves along ($\delta$,  -$\delta$) and ($\delta$,  0) directions around $\bm{q}$ = (4, 1) measured at various temperatures.}
\label{fig:5}
\end{figure}
%====================================================================

To elucidate the relationship between phonon and superconductivity in the BiS$_2$-based superconductors, we examined the phonon dispersion by observing a thermal diffuse scattering (TDS), a diffuse scattering process by lattice thermal vibration. For a phonon-mediated conventional superconductor, there can be a phonon frequency softening and/or line-width broadening at a momentum transfer $\bm{q}$ = 2$\bm{k}_F$ across the superconducting phase transition \cite{Kohn1, Kohn2, Kohn3, Kohn4, Kohn5, Kohn5, Kohn6}. 

To the 1st order, the intensity of TDS by unpolarized x-ray at a momentum transfer $\bm{q}$ can be represented as \cite{TDS}
%====================================================================
\begin{equation}
\begin{split}
    I(\bm{q}) \propto  & \sum_{j = 1}^{3n}\frac{1}{\omega_{\bm{q},j}}\coth(\frac{\hbar\omega_{\bm{q},j}}{2k_{B}T}) \\
    & \times \Big|\sum_{s = 1}^{n}\frac{f_s}{\sqrt{\mu_s}}e^{-M_s}(\bm{q}\cdot\bm{e}_{\bm{q},j,s})e^{-i\bm{q}\cdot\tau_{s}}\Big|^2 ,
\label{eq:eq2}
\end{split}
\end{equation}
%====================================================================
where the subscripts $s$ and $j$ represent the atom in a unit cell and the phonon modes, respectively. Here, $\omega$, $\bm{e}$, $f$, $\mu$, $e^{-M}$, and $\tau$ refer to phonon frequency, eigenvector for the phonon mode, atomic form factor, atomic mass, the Debye-Waller factor, and basis vector in real space, respectively. From this equation, we can tell that the main contribution to the TDS intensity will be from the low lying phonon modes, where significant electron-phonon couplings have been estimated in BiS$_2$-based superconductors. In addition, from the $\coth$-dependence with $1/T$, we expect the thermal change of intensity will be negligible in the small temperature range around T$_c$.

Figure 3 (a) shows some representative intensity maps in the (H, K, 0) plane measured at 4.3, 5.5, 8.0, and 10 K. We were not able to observe any meaningful $\bm{q}$-dependent or T-dependent changes in the TDS. For more careful inspection, we plot intensity curves along several directions, for example, along diagonal ($\delta$,  -$\delta$) and  horizontal ($\delta$,  0) directions in Fig. 3 (b). No clear and consistent change is found across T$_c$ while some anomalous kink can be attributed to the 3 x 3 phase. It should be noted that TDS intensity is an energy-integrated measurement and it is always possible that a small thermal population at low temperature and/or not-enough x-ray flux may hinder the observation of a small phonon anomaly. With these limitations in mind, we conclude it is likely that the the change of phonon frequency across the superconducting phase transition is small to be detected.

\section{DISCUSSION}

Here we report that NdO$_{1-x}$F$_{x}$BiS$_2$ (x = 0.37 and 0.3) develops a CDW, a long-sought structural ground state in the BiS$_2$-based superconductors that coexist with superconductivity. The origin of a CDW is an interesting and often challenging question. It can be from FSN as in simple Pierels system, $\bm{q}$-dependent electron-phonon coupling as in some TMDC, or electron-electron correlation in cuprates \cite{Plummer}. While the \textbf{Q}$_{\rm{CDW}}$ seems to be far from the FSN vectors known so far, electron-phonon coupling calculations were not successful to predict the wavevector yet. More careful studies considering the interlayer couplings and even electronic correlations could be required.   

The discrepancy between the calculations and measurements might be from the structural instability to the monoclinic phase in the BiS$_2$-based system. At high pressure, LaO$_{1-x}$F$_x$BiS$_2$ undergoes a structural phase transition to P2$_1$/m and yields higher T$_c$ \cite{HPEnhance}. According to the first-principle calculations, in the monoclinic phase, the magnitude of the interlayer hopping becomes much larger stabilizing the electronic state \cite{Bilayer, SuzukiRev}. Our intensity map at (H, K, 0) in the Supplementary show that the x-ray reflections no longer satisfy the P4/nmm extinction conditions, probably due to a monoclinic phase transition as observed in some other BiS$_2$-based systems \cite{Monoclinic}. The enhanced interlayer hopping in the monoclinic structure, therefore, can be responsible for the L-dependence of our CDW wavevector. We claim that the interlayer coupling overlooked so far in the phonon and band calculations can be essential to properly describe the electron-phonon coupling, FSN, CDW, and the nature of superconductivity in BiS$_2$-based superconductors. 

Lastly, the role of CDW in superconductivity has been an active research subject since Fr\"{o}hlich tried to explain superconductivity with sliding of CDW, a concept conceived even before BCS formalism \cite{Frohlich}. While there is accumulating evidence that a CDW could be a ubiquitous phase in unconventional superconductors \cite{Dean, Kis}, the relation with superconductivity remains elusive. Here, we suggested that the charge density wave and superconductivity might be in competition while coexisting over a certain doping range in the phase diagram. Like lattice vibrations in a conventional superconductor, charge/orbital fluctuation near the quantum critical point is argued to be capable of mediating superconductivity. Considering the absence of momentum- and temperature- dependent anomalies in the phonon spectrum across T$_c$, we cannot exclude the unconventional pairing mechanism. The study on the interplay between CDW and superconductivity in BiS$_2$-based superconductors, by changing external perturbations like doping or pressure will be a fascinating topic that requires lots of future attention.

\section{CONCLUSION}

In summary, we found an intrinsic and long-range CDW phase in a NdO$_{1-x}$F$_{x}$BiS$_2$ (x = 0.37 and 0.3) superconductor by x-ray diffraction measurements. The CDW wavevector is \textbf{Q}$_{\rm{CDW}}$ $\approx$ (0.17, 0.17, 0.5) with clear L-dependence and the atomic displacements are transverse and in diagonal directions. The wavevector does not match theoretical estimations from phonon softening or FSN and therefore places an important constraint on forthcoming studies. We measured TDS across the T$_c$ and could not find meaningful $\bm{q}$- and T- dependent changes.
%, in favor of unconventional superconductivity in BiS$_2$-based superconductors
\section{ACKNOWLEDGMENTS}

We thanks D. Louca and C. Park for fruitful discussions. Cornell High Energy Synchrotron Source was supported by the NSF and NIH/National Institute of General Medical Sciences via NSF award no. DMR-1332208.

%%=======================BIBLIOGRAPHY==============================

\end{document}